\documentclass[final,5p,times,twocolumn,numbers]{elsarticle}

\usepackage{amssymb}
\usepackage{amsmath}
\usepackage{lipsum}
\usepackage{booktabs}
\usepackage{slashed}
\usepackage[compat=1.1.0]{tikz-feynman}
\usepackage{subcaption}
\usepackage{xurl}
\usepackage[hidelinks]{hyperref}

\journal{Physics Letters B}

\begin{document}

\begin{frontmatter}



\title{Exploring X17 Couplings through Meson Dalitz Decays}

\author[hit]{Xuan Wang}
\ead{25S011011@stu.hit.edu.cn}

\author[hit]{Tianhong Wang}
\ead{thwang@hit.edu.cn}

\author[hit]{Si-Qi Zhang}
\ead{26b911024@stu.hit.edu.cn}

\author[hbu]{Guo-Li Wang}
\ead{wgl@hbu.edu.cn}

\affiliation[hit]{
  organization={School of Physics, Harbin Institute of Technology},
  city={Harbin},
  postcode={150001},
  country={China}
}

\affiliation[hbu]{
  organization={Department of Physics, Hebei University},
  city={Baoding},
  postcode={071002},
  country={China}
}



\begin{abstract}
The ATOMKI collaboration has reported anomalous $e^+e^-$ pair production in a series of nuclear transitions, which has been interpreted as a possible signature of a new light boson, X17. Meson Dalitz decays provide a complementary probe of the X17 interaction structure. However, the widely studied vector coupling scenario is in tension with the combined constraints from existing $D^{*0}\to D^0e^+e^-$, $D_s^{*+}\to D_s^+e^+e^-$ measurements and nuclear transition analyses. In this work, we investigate the X17 contributions to meson Dalitz decays within a framework incorporating both vector and axial-vector couplings, employing the Bethe--Salpeter approach to calculate the hadronic transition matrix elements. With $u$ and $d$ quark couplings taken from previous analyses, we identify allowed regions for the charm and strange quark couplings, which require the axial-vector coupling strength $|g_{c,s}^A|\sim10^{-3}$. We obtain four degenerate best-fit solutions, one of which is $(g_c^V,g_c^A,g_s^V,g_s^A)=(8.32,1.09,-2.62,1.36)\times10^{-3}$, and estimate the decay ratio $R^{\mathrm{th}}(D^{*+}\to D^+e^+e^-)$, which significantly exceeds the Standard Model prediction.
\end{abstract}







\end{frontmatter}




\section{Introduction}
\label{sec:introduction}

The search for physics beyond the Standard Model remains a focus of modern particle physics. In 2016, the ATOMKI Collaboration reported an anomalous excess of electron--positron pairs produced in the de-excitation of an excited state of $^{8}\mathrm{Be}$~\cite{Krasznahorkay2016Be}. The excess, observed at an opening angle of approximately $\theta \simeq 140^\circ$, was reported with a statistical significance of $6.8\sigma$ relative to the expected internal pair creation (IPC) background. This anomaly was interpreted as a possible signature of a hypothetical neutral boson with a mass of $m_X=(16.70\pm0.35_{\mathrm{stat}}\pm0.50_{\mathrm{syst}})\,\mathrm{MeV}$. Subsequently, the ATOMKI Collaboration reported similar anomalous structures in nuclear transitions involving $^{4}\mathrm{He}$~\cite{Krasznahorkay2021He} and $^{12}\mathrm{C}$~\cite{Krasznahorkay2022C}, with inferred masses consistently around $17\,\mathrm{MeV}$. The hypothetical particle associated with these observations is commonly referred to as X17.

This anomaly has attracted considerable attention. The NA64 Collaboration searched for X17 production through $e^-Z\to e^-ZX$ but observed no significant signal~\cite{NA642018}. In 2024, an independent experiment reported an excess exceeding $4\sigma$ in the $^{7}\mathrm{Li}(p,e^+e^-){}^{8}\mathrm{Be}$ reaction~\cite{TranTheAnh2024}, consistent with the ATOMKI observations. The MEG II Collaboration also investigated the same reaction but found no significant evidence for X17, setting upper limits on its production~\cite{MEGII2025}. The PADME Collaboration searched for X17 through $e^+e^-\to X\to e^+e^-$ and reported a weak excess near $16.90\,\mathrm{MeV}$ with a local significance of approximately $2.5\sigma$~\cite{PADME2025}. Recently, the BESIII Collaboration searched for X17 in the charmonium transitions $\chi_{cJ}\to J/\psi X$ but observed no significant signal, setting an upper limit of $|\varepsilon_c|<1.2\times10^{-2}$ at the $90\%$ confidence level on the X17 coupling to charm quarks~\cite{BESIII2026X17}.

Since the spin and parity of X17 remain undetermined, various theoretical interpretations have been proposed, among which the light spin-1 boson scenarios receive particular attention. A protophobic vector boson has been proposed to explain the original $^{8}\mathrm{Be}$ anomaly~\cite{Feng2016}, in which the effective proton coupling, $\epsilon_p=2\epsilon_u+\epsilon_d$, is strongly constrained by the NA48/2 experiment~\cite{NA482015}. The axial-vector interpretation has also been explored~\cite{Kozaczuk2017}. However, rare pion decay measurements by SINDRUM-I~\cite{SINDRUM1989} place strong constraints on simple vector and axial-vector explanations, unless the quark and lepton couplings are arranged to suppress the decay amplitude~\cite{Hostert2023}.  A recent systematic analysis of X17 with both vector and axial-vector couplings identified parameter regions compatible with the ATOMKI observations when the $^{12}\mathrm{C}$ data were excluded~\cite{Fieg2026}. Further theoretical studies and reviews exploring the anomaly from various perspectives can be found in Refs.~\cite{ZhangMiller2021,Mommers2024,DelleRose2017,shedding}.

Meson Dalitz decays, such as $V\to Pe^+e^-$, provide a sensitive probe of possible X17 contributions. The $D_{(s)}^*\to D_{(s)}e^+e^-$ and $B_{(s)}^*\to B_{(s)}e^+e^-$ channels have been studied within heavy quark effective theory (HQET) combined with vector meson dominance (VMD)~\cite{Castro2021}, and covariant confined quark model (CCQM)~\cite{Tran2025}, assuming generation universal couplings ($\varepsilon_u=\varepsilon_c=\varepsilon_t$, $\varepsilon_d=\varepsilon_s=\varepsilon_b$). A subsequent study relaxed this assumption and fitted the coupling parameters to existing experimental results, but obtained two solutions for $|\varepsilon_u|$ near $6.1\times10^{-2}$ and $6.7\times10^{-2}$~\cite{Lee2026}, both of which are in tension with the ATOMKI measurements.

In this work, we first calculate the X17 contributions to $D^{*0}\to D^0e^+e^-$ and $D_s^{*+}\to D_s^+e^+e^-$ in the vector couping scenario and compare our results with previous studies. We find the same difficulty: assuming generation universal couplings, the X17 contribution is too small in the $D^{*0}$ channel but too large in the $D_s^{*+}$ channel to account for the deviations between the experimental results and the Standard Model predictions. We therefore extend the model to include both vector and axial-vector couplings, aiming to accommodate different measurements. The hadronic transition matrix elements are calculated using the Bethe--Salpeter (BS) method. Taking the $(g_u^V,g_d^V)$ from nuclear transition analyses, we identify allowed regions in the $(g_c^V,g_s^V,g_c^A,g_s^A)$ parameter space and perform a combined four-channel $\chi^2$ fit. We further predict the possible X17 contribution to $D^{*+}\to D^+e^+e^-$ as a sensitive test of our theoretical framework.

The paper is organized as follows. Sec. 2 presents the X17 effective interactions and the formalism for Dalitz decays, followed by a brief introduction to the BS method. Numerical results are discussed in Sec. 3. Finally, Sec. 4 gives the main conclusions and outlook.

\section{Effective interaction Lagrangian and Dalitz decay}
\label{sec:method}
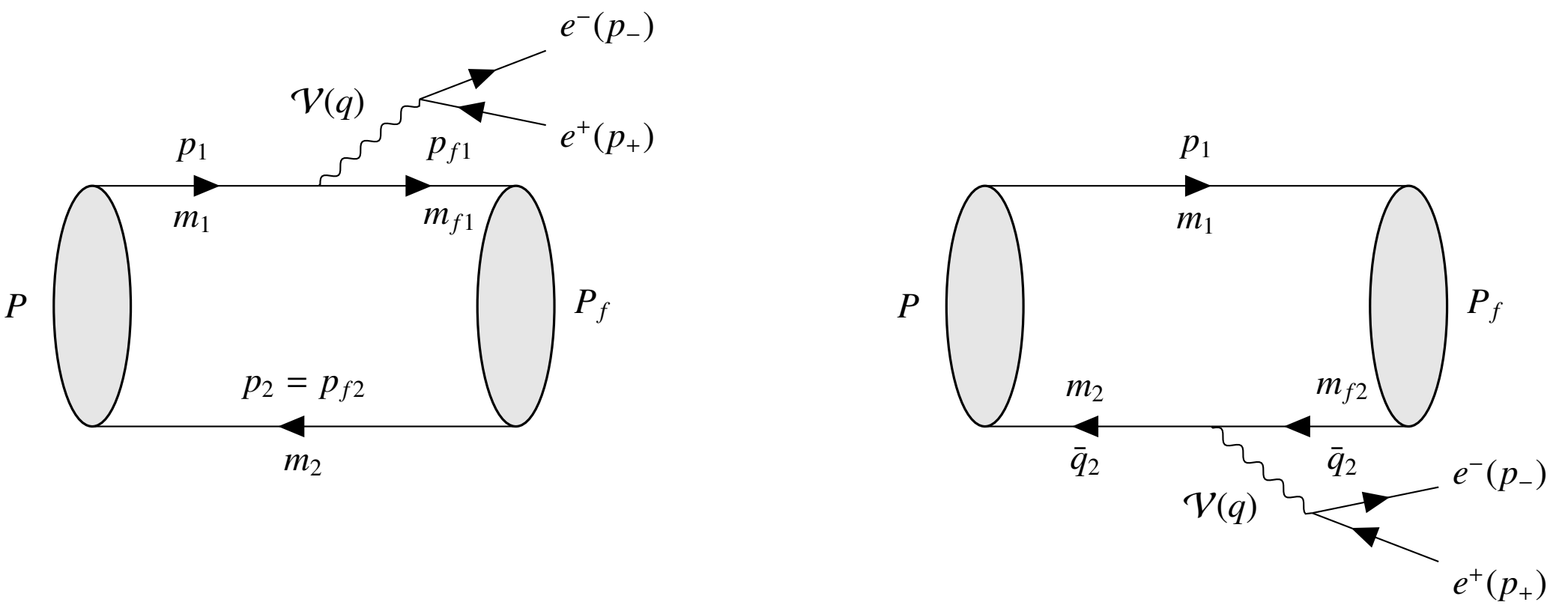
\begin{figure*}[t]
\centering

\begin{subfigure}[t]{0.4\textwidth}
\centering

\begin{tikzpicture}[
  scale=0.85,
  every node/.style={font=\small}
]

\path[use as bounding box]
    (-0.55,-3.00) rectangle (5.85,3.00);

\begin{feynman}

\vertex (a) at (0,1.25);
\vertex (b) at (0,-1.25);

\vertex (c) at (2.35,1.25);

\vertex (d) at (4.40,1.25);
\vertex (e) at (4.40,-1.25);

\vertex (x) at (3.40,2.15);

\vertex (em) at (5.35,2.90) {$e^-(p_-)$};
\vertex (ep) at (5.35,1.75) {$e^+(p_+)$};

\diagram*{
    (a) -- [fermion] (c),
    (c) -- [fermion] (d),

    (b) -- [anti fermion] (e),

    (c) -- [boson] (x),

    (x) -- [fermion] (em),
    (x) -- [anti fermion] (ep),
};

\end{feynman}

\draw[fill=gray!20, draw=black, line width=0.6pt]
    (0,0) ellipse (0.40cm and 1.25cm);

\draw[fill=gray!20, draw=black, line width=0.6pt]
    (4.40,0) ellipse (0.40cm and 1.25cm);

\node at (-0.80,0) {$P$};
\node at (5.20,0) {$P_f$};

\node at (1.05,1.62) {$p_1$};
\node at (1.05,0.88) {$m_1$};

\node at (3.72,1.62) {$p_{f1}$};
\node at (3.72,0.88) {$m_{f1}$};

\node at (2.20,-1.65) {$m_2$};
\node at (2.20,-0.86) {$p_2=p_{f2}$};

\node at (2.45,2.10) {$\mathcal{V}(q)$};

\end{tikzpicture}

\caption{}
\label{fig:quark-emission}

\end{subfigure}
\hspace{0.02\textwidth}
\begin{subfigure}[t]{0.4\textwidth}
\centering

\begin{tikzpicture}[
  scale=0.85,
  every node/.style={font=\small}
]

\path[use as bounding box]
    (-0.55,-3.00) rectangle (5.85,3.00);

\begin{feynman}

\vertex (a) at (0,1.25);
\vertex (b) at (0,-1.25);

\vertex (c) at (2.35,-1.25);

\vertex (d) at (4.40,1.25);
\vertex (e) at (4.40,-1.25);

\vertex (x) at (3.40,-2.15);

\vertex (em) at (5.35,-1.75) {$e^-(p_-)$};
\vertex (ep) at (5.35,-2.90) {$e^+(p_+)$};

\diagram*{
    (a) -- [fermion] (d),

    (b) -- [anti fermion] (c),
    (c) -- [anti fermion] (e),

    (c) -- [boson] (x),

    (x) -- [fermion] (em),
    (x) -- [anti fermion] (ep),
};

\end{feynman}

\draw[fill=gray!20, draw=black, line width=0.6pt]
    (0,0) ellipse (0.40cm and 1.25cm);

\draw[fill=gray!20, draw=black, line width=0.6pt]
    (4.40,0) ellipse (0.40cm and 1.25cm);

\node at (-0.80,0) {$P$};
\node at (5.20,0) {$P_f$};

\node at (2.20,1.65) {$p_1$};
\node at (2.20,0.86) {$m_1$};

\node at (1.05,-1.62) {$\bar q_2$};
\node at (1.05,-0.88) {$m_2$};

\node at (3.72,-1.62) {$\bar q_2$};
\node at (3.72,-0.88) {$m_{f2}$};

\node at (2.45,-2.10) {$\mathcal{V}(q)$};

\end{tikzpicture}

\caption{}
\label{fig:antiquark-emission}

\end{subfigure}

\caption{
Feynman diagrams for the electromagnetic and X17 contributions to
$V(P)\rightarrow P(P_f)e^+e^-$. 
The intermediate boson $\mathcal{V}$ is emitted from the quark line (a)
or antiquark line (b), with $\mathcal{V}=\gamma^*$ or X17.
}
\label{fig:feynman}

\end{figure*}

We consider X17 as a massive spin-1 boson that interacts with Standard Model fermions through both vector and axial-vector couplings. The effective interaction Lagrangian is given by
\begin{align}
\mathcal{L}_{\mathrm{int}}
= e X_\mu \sum_f \bar{f}\gamma^\mu
\left(g_f^V+g_f^A\gamma_5\right)f,
\label{Lagrangian}
\end{align}
where $X_\mu$ denotes the X17 field, $e$ is the elementary electric charge, and $g_f^V$ and $g_f^A$ are the dimensionless vector and axial-vector coupling constants, respectively.

The Dalitz decay $V(P,\epsilon)\to P(P_f)e^+(p_+)e^-(p_-)$ proceeds through a virtual photon in the Standard Model, while the proposed X17 boson introduces an additional contribution, as shown in Fig.~\ref{fig:feynman}. The invariant amplitude can be written as
\begin{align}
i\mathcal{M}_{\gamma^*}
&=e H_{\gamma^*}^\mu
\frac{-ig_{\mu\nu}}{q^2+i\varepsilon}
\bar{u}(p_-)(-ie\gamma^\nu)\text{\textit{v}}(p_+),
\nonumber\\
i\mathcal{M}_X
&=e H_X^\mu
\frac{-i(g_{\mu\nu}-q_\mu q_\nu/m_X^2)}
{q^2-m_X^2+im_X\Gamma_X}
\nonumber\\
&\quad\times
\bar{u}(p_-)
(-ie\gamma^\nu)
\left(g_e^V+g_e^A\gamma_5\right)
\text{\textit{v}}(p_+).
\end{align}
Here $q=p_++p_-=P-P_f$ is the momentum transferred to the lepton pair, $m_X$ and $\Gamma_X$ denote the mass and total decay width of X17, respectively. We use $m_X=17\,\mathrm{MeV}$ throughout this work. $H_{\gamma^*}^\mu$ and $H_X^\mu$ are the corresponding hadronic transition matrix elements, defined as
\begin{align}
H_{\mathrm{\gamma^*}}^\mu
&=\left\langle P\left|
\sum_q Q_q\bar q\gamma^\mu q
\right|V\right\rangle,
\nonumber\\
H_X^\mu
&=\left\langle P\left|
\sum_q\bar q\gamma^\mu
(g_q^V+g_q^A\gamma_5)q
\right|V\right\rangle,
\label{eq:hadronic_currents}
\end{align}
where $Q_q$ denotes the electric charge of quark $q$ in units of $e$. 

The  decay width is
\begin{align}
\Gamma_{ee}
&=\frac{1}{512\pi^3 M^3}
\int\frac{dq^2}{q^2}\lambda^{1/2}(M^2,q^2,M_f^2)
\nonumber\\
&\quad\times\lambda^{1/2}(q^2,m_e^2,m_e^2)
\int d\cos\theta\,
\overline{|\mathcal M_{\gamma^*}+\mathcal M_{X}|^2}.
\label{eq:SM_three_body_width}
\end{align}
Here, $M$ and $M_f$ are the masses of the initial and final mesons, respectively, $\lambda(a,b,c)=a^2+b^2+c^2-2ab-2ac-2bc$, and $\theta$ is the angle between the electron momentum (in the $e^+e^-$ center-of-mass frame) and the final state meson momentum.

For the X17 contribution, we employ the narrow width approximation (NWA),
\begin{align}
\frac{1}{(q^2-m_X^2)^2+m_X^2\Gamma_X^2}
&\simeq\frac{\pi}{m_X\Gamma_X}
\delta(q^2-m_X^2).
\label{eq:NWA}
\end{align}
Therefore, the interference between the X17 and virtual photon contributions can be neglected, and the total decay width $\Gamma_{ee}$ can be approximated as the sum of $\Gamma^{\gamma^*}_{ee}$ and $\Gamma^X_{ee}$. Assuming that $\Gamma_X\simeq\Gamma(X\to e^+e^-)$, the electron coupling parameter cancels out in the final result. Performing the integration yields a quadratic form
\begin{align}
\Gamma_{ee}^{X}
&=C_{V,11}(g_{q_1}^V)^2
+C_{V,12}g_{q_1}^Vg_{q_2}^V
+C_{V,22}(g_{q_2}^V)^2
\nonumber\\
&\quad+C_{A,11}(g_{q_1}^A)^2
+C_{A,12}g_{q_1}^Ag_{q_2}^A
+C_{A,22}(g_{q_2}^A)^2.
\label{eq:quadratic_width}
\end{align}
Here $q_1$ and $q_2$ denote the constituent quarks, and the coefficients $C_{V,ij}$ and $C_{A,ij}$ are obtained by evaluating the hadronic matrix elements.

To compare with experimental results, we use the ratio of the Dalitz decay width to the corresponding radiative decay width,
\begin{align}
R&=\frac{\Gamma(V\to Pe^+e^-)}
{\Gamma(V\to P\gamma)},
\label{eq:R_definition}
\end{align}
and
\begin{align}
R&=R^{\mathrm{SM}}+R^X.
\label{eq:R_theory}
\end{align}

In this work, we employ the instantaneous BS method to calculate the hadronic transition matrix elements. For the initial vector meson with $J^P=1^-$, the relativistic Salpeter wave function takes the form~\cite{WangHeavyVector2006}
\begin{align}
\varphi_{1^-}(k_\perp)
={}&(k_\perp\cdot\epsilon)
\Bigg[f_1+\frac{\slashed P}{M}f_2
+\frac{\slashed k_\perp}{M}f_3
+\frac{\slashed P\slashed k_\perp}{M^2}f_4\Bigg]
\nonumber\\
&+M\slashed\epsilon
\Bigg[f_5+\frac{\slashed P}{M}f_6
+\frac{\slashed k_\perp}{M}f_7
+\frac{\slashed P\slashed k_\perp}{M^2}f_8\Bigg],
\label{eq:BS_vector}
\end{align}
where $P$ and $k$ denote the total and relative momenta of the initial meson, respectively, $\epsilon^\mu$ is the polarization vector, and $k_\perp^\mu=k^\mu-(P\cdot k)P^\mu/M^2$. The $f_i$ are scalar functions of $k_\perp$.

For the final pseudoscalar meson with $J^P=0^-$, the wave function is~\cite{KimWangKinetic2004}
\begin{equation}
\varphi_{0^-}(k_{f\perp})
=M_f\left[
g_1+\frac{\slashed P_f}{M_f}g_2
+\frac{\slashed k_{f\perp}}{M_f}g_3
+\frac{\slashed P_f\slashed k_{f\perp}}{M_f^2}g_4
\right]\gamma_5.
\label{eq:BS_pseudoscalar}
\end{equation}
Here, $P_f$ and $k_f$ denote the total and relative momenta of the final meson, respectively, with $k_{f\perp}^\mu=k_f^\mu-(P_f\cdot k_f)P_f^\mu/M_f^2$, while the $g_i$ are scalar functions of $k_{f\perp}$.

Within the Mandelstam formalism, the hadronic transition matrix element is expressed as an overlap integral of the initial and final Salpeter wave functions. The two contributions associated with the quark and antiquark currents shown in Fig.~\ref{fig:feynman} are retained as~\cite{WangBc2S2016}
\begin{align}
H^\mu
&=\int\frac{d^3\mathbf{k}}{(2\pi)^3}
\operatorname{Tr}\Bigg[
\frac{\slashed P}{M}
\overline{\varphi}_{0^-}^{++}(k_{\perp}+\alpha_2P_{f\perp})
\Gamma_1^\mu\varphi_{1^-}^{++}(k_\perp)
\nonumber\\
&\qquad-
\overline{\varphi}_{0^-}^{++}(k_{\perp}-\alpha_1P_{f\perp})
\frac{\slashed P}{M}
\varphi_{1^-}^{++}(k_\perp)\Gamma_2^\mu
\Bigg].
\label{eq:BS_overlap}
\end{align}
Here, $\overline{\varphi}^{++}$ is the positive energy component of the wave function, $\alpha_{1,2}=m_{1,2}/(m_1+m_2)$, and $P_{f\perp}^\mu=P_f^\mu-(P\cdot P_f) P^\mu/M^2$. The electromagnetic and X17-quark interaction vertices are given by
\begin{align}
\Gamma_{i,\gamma}^\mu&=Q_{q_i}\gamma^\mu,\nonumber\\
\Gamma_{i,X}^\mu&=\gamma^\mu(g_{q_i}^V+g_{q_i}^A\gamma_5),
\quad i=1,2,
\label{eq:BS_vertices}
\end{align}
where $Q_{q_i}$ and $g_{q_i}^{V,A}$ follow the definitions in Eqs.~\eqref{eq:hadronic_currents} and \eqref{Lagrangian}, respectively.

\section{Discussion}
\label{result}
In our calculation, we adopt three benchmarks from Ref.~\cite{Feng2016,DentonGehrlein2023}. We first consider the scenario without including axial-vector couplings and assume $g_c^V=g_u^V$ and $g_s^V=g_d^V$. The results are listed in Table~\ref{tab:pure_vector}. For all three benchmarks, the X17 contribution is too small in the $D^{*0}$ channel but too large in the $D_s^{*+}$ channel to account for the deviations between the experimental measurements and the Standard Model predictions. We further attempted to identify a common allowed parameter region but found none consistent with the nuclear constraints. These results agree with the tension reported by Lee et al.~\cite{Lee2026} and motivate us to consider a more general X17 interaction structure.

\begin{table*}[t]
\caption{\label{tab:pure_vector}
Calculated Dalitz decay ratios for the vector coupling scenario. Benchmarks I, II, and III adopt
$(g_u^V,g_d^V)=(3.7,-7.4)$, $(0.5,-2.9)$, and
$(0.9,-2.5)$, respectively, in units of $10^{-3}$.
We assume $g_c^V=g_u^V$, $g_s^V=g_d^V$, and $g_q^A=0$.
All displayed ratios are in units of $10^{-3}$.
}
\renewcommand{\arraystretch}{1.5}
\begin{tabular*}{\textwidth}{@{\extracolsep{\fill}}lcccccccc}\toprule
& & & \multicolumn{2}{c}{Benchmark I}
& \multicolumn{2}{c}{Benchmark II}
& \multicolumn{2}{c}{Benchmark III}\\
Channel & $R^{\rm SM}$ & $R^{\rm exp}$
& $R^X$ & $R^{\rm th}$
& $R^X$ & $R^{\rm th}$
& $R^X$ & $R^{\rm th}$\\
\midrule$D^{*0}\to D^0e^+e^-$& 6.4 & $11.08\pm0.90$~\cite{BESIII2021Dstar0}
& $3.0\times10^{-2}$ & 6.5
& $5.5\times10^{-4}$ & 6.4
& $1.8\times10^{-3}$ & 6.4\\

$D_s^{*+}\to D_s^+e^+e^-$
& 6.5 & $7.78\pm0.73$~\cite{BESIII2026Ds}
& 31 & 38
& 6.8 & 13
& 4.2 & 11\\

$D^{*+}\to D^+e^+e^-$& 6.4& $-$
& 4.6 & 11
& 0.93 & 7.4
& 0.59 & 7.0\\

$\psi(2S)\to\eta_c e^+e^-$& 8.9 & $11.10\pm2.10$~\cite{BESIII2022Psi}
& $3.1\times10^{-2}$ & 9.0
& $5.6\times10^{-4}$ & 9.0
& $1.8\times10^{-3}$ & 9.0\\
\bottomrule
\end{tabular*}
\end{table*}

\begin{figure*}[!t]
    \centering
    \includegraphics[width=0.85\textwidth]{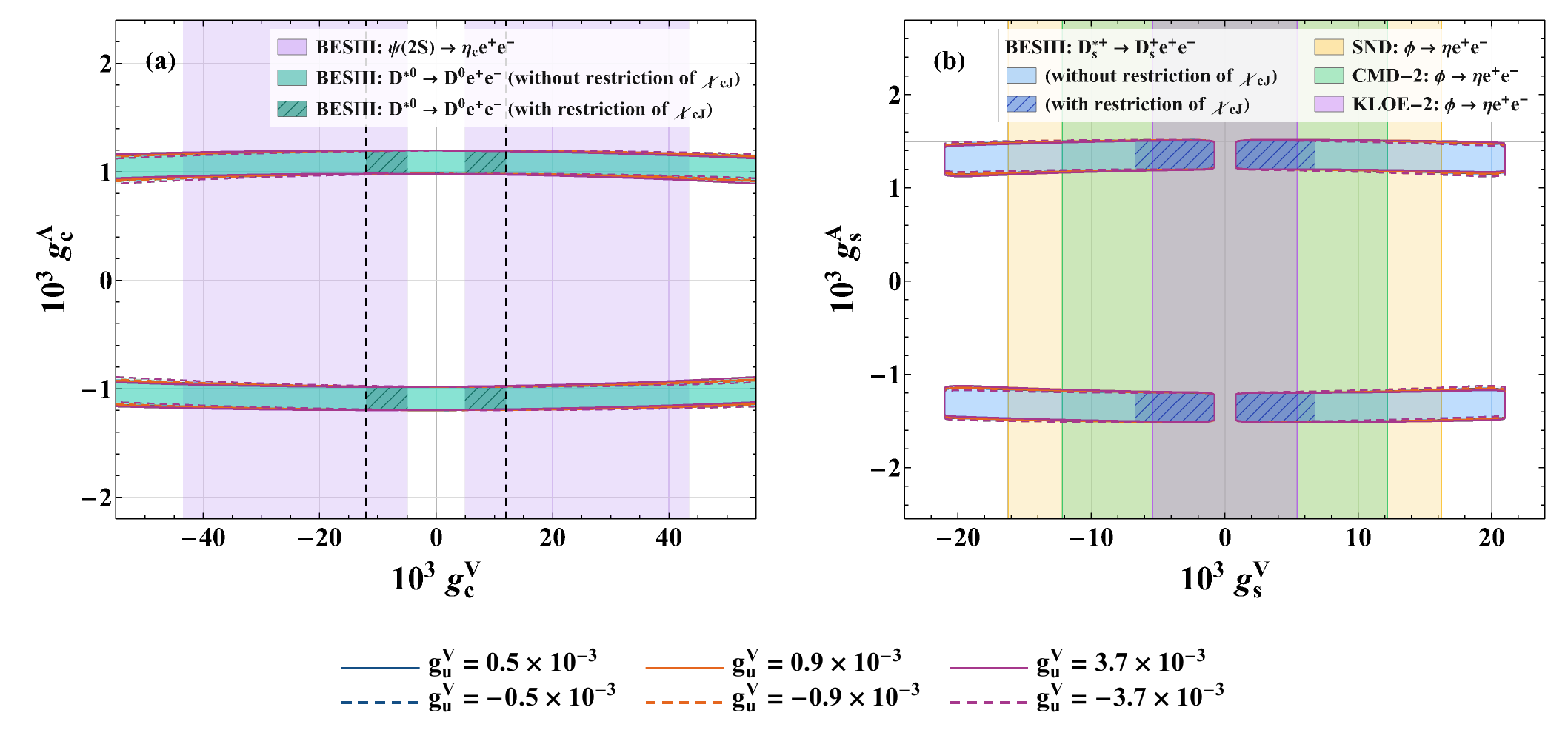}
    \caption{Allowed coupling regions at the experimental $1\sigma$ level:
    (a) $(g_c^V,g_c^A)$ from the $D^{*0}$ and $\psi(2S)$ channels;
    (b) $(g_s^V,g_s^A)$ regions from $D_s^{*+}$,
    together with the independent $\phi$ constraints.
    The boundary curves correspond to
    $|g_u^V|=(0.5,0.9,3.7)\times10^{-3}$,
    with solid (dashed) curves indicating positive (negative) $g_u^V$. The vertical dashed lines in panel (a) represent the restriction from Ref.~\cite{BESIII2026X17}.}
    \label{fig:coupling_allowed}
\end{figure*}

To simplify the model, we assume $g_u^A=g_d^A=0$ while treating the vector and axial-vector couplings of the $c$ and $s$ quarks as independent parameters. For each benchmark in Table~\ref{tab:pure_vector}, we constrain the charm quark couplings using the $D^{*0}\to D^0e^+e^-$ and $\psi(2S)\to\eta_c e^+e^-$ measurements. We impose the $1\sigma$ bounds,
\begin{align}
R_i^{\mathrm{exp}}-\sigma_i
\leq R_i^{\mathrm{SM}}+R_i^X
\leq R_i^{\mathrm{exp}}+\sigma_i,
\label{eq:allowed_couplings}
\end{align}
where $i$ labels the decay channel. The resulting charm quark coupling regions are shown in Fig.~\ref{fig:coupling_allowed}(a), determined by the overlap of the two experimental constraints. Since the nuclear constraints fix only the relative sign of the $u$ and $d$ vector couplings, we consider both signs of $g_u^V$, represented by solid and dashed curves, respectively. Reversing the sign of $g_u^V$ reflects the allowed regions about $g_c^V=0$, while the regions corresponding to the three benchmarks nearly overlap for each sign. The value of $g_c^V$ can vary over a wide range, whereas $|g_c^A|\simeq1.1\times10^{-3}$. In addition, we also include the restriction $|g_c^V|<1.2\times10^{-2}$ from Ref.~\cite{BESIII2026X17}, with the darker shaded regions indicating the allowed parameter space after imposing this constraint.

Adopting the allowed charm quark couplings in Fig.~\ref{fig:coupling_allowed}(a) as input and imposing the $D_s^{*+}\to D_s^+e^+e^-$ constraint, we obtain the maximal allowed regions in the $(g_s^V,g_s^A)$ plane. As shown in Fig.~\ref{fig:coupling_allowed}(b), we compare these regions with the constraints on $g_s^V$ obtained by Lee et al.~\cite{Lee2026} from the SND~\cite{SND2001}, CMD-2~\cite{CMD22001}, and KLOE-2~\cite{KLOE2015} measurements of $\phi\to\eta e^+e^-$. The upper and lower bands correspond the positive and negative $g_c^A$ branches in Fig.~\ref{fig:coupling_allowed}(a), respectively. Notably, the allowed regions exclude $g_s^A=0$ and require $|g_s^A|\gtrsim1.12\times10^{-3}$. The negative axial-vector interference coefficient $C_{A,12}^{D_s^{*+}}$ results in destructive interference between the $c$ and $s$ contributions, minimizing the axial-vector contribution near $g_s^A\simeq1.24g_c^A$. Thus, a nonzero $g_s^A$ can suppress the X17 contribution in the $D_s^{*+}$ channel while the corresponding $g_c^A$ enhances it in the $D^{*0}$ channel.

\begin{figure*}[!t]
    \centering
    \includegraphics[width=0.85\textwidth]{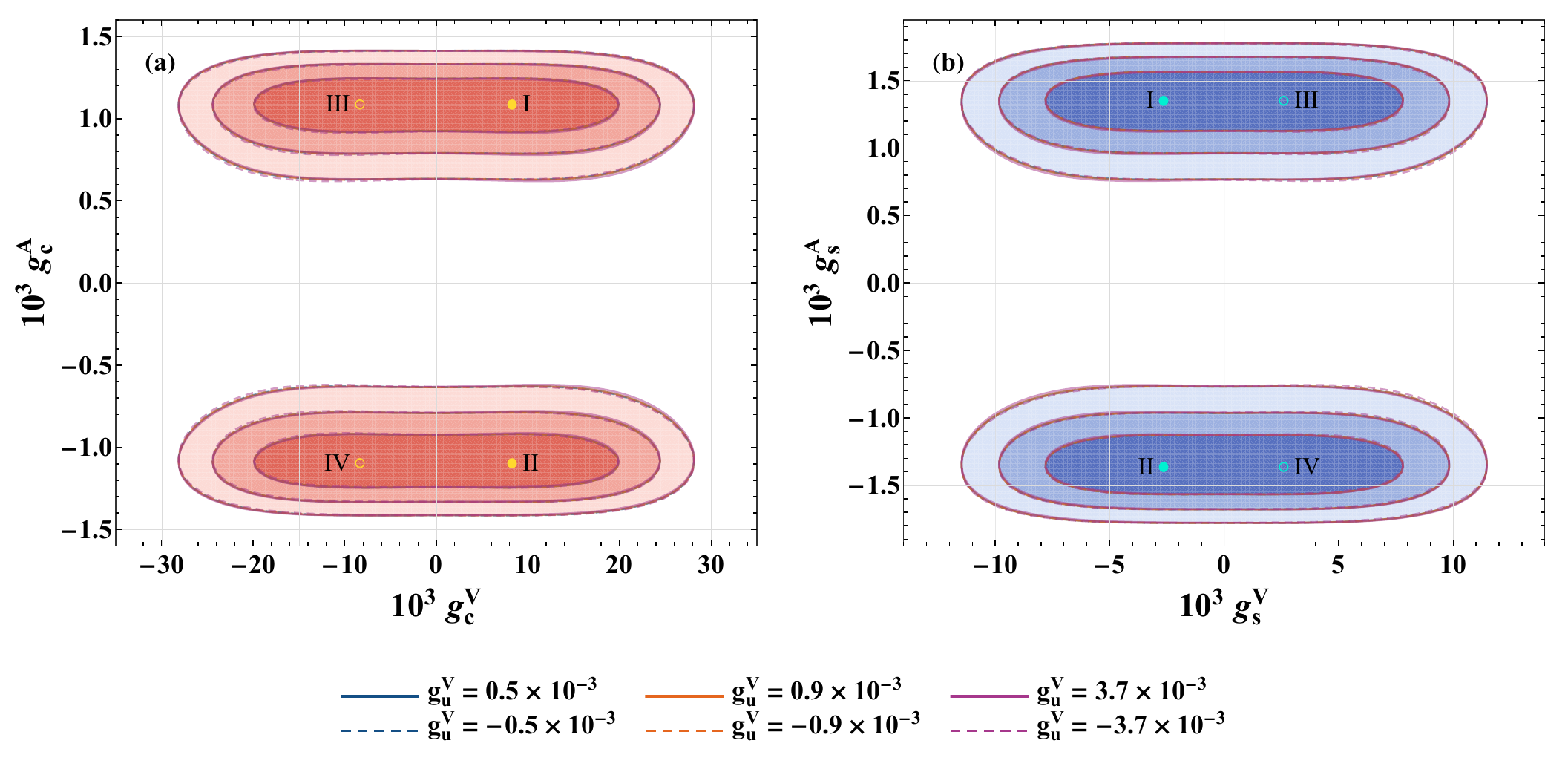}
    \caption{Best-fit points and $1\sigma$, $2\sigma$, and $3\sigma$ regions for the three benchmarks in the (a) $(g_c^V,g_c^A)$ and (b) $(g_s^V,g_s^A)$ planes. Filled markers ($\mathrm{I}$, $\mathrm{II}$) and open markers ($\mathrm{III}$, $\mathrm{IV}$) correspond to positive and negative $g_u^V$, respectively.}
    \label{fig:fit}
\end{figure*}

To further assess the compatibility of the four decay channels, we perform a combined $\chi^2$ fit for each of the three nuclear benchmarks. With $g_u^V$ fixed, the free parameters are $\theta=(g_c^V,g_c^A,g_s^V,g_s^A)$. We define
\begin{align}
\chi^2(\theta)=\sum_i\left[\frac{R_i^{\mathrm{th}}(\theta)-R_i^{\mathrm{exp}}}{\sigma_i}\right]^2,
\label{eq:chi2}
\end{align}
where $i$ runs over the $D^{*0}$, $D_s^{*+}$, $\psi(2S)$, and $\phi$ channels, and $\sigma_i$ denotes the experimental uncertainty. The first three theoretical ratios are calculated using the BS approach, while for the $\phi$ channel we adopt the VMD approximation $R_\phi^{\mathrm{th}}=8.30\times10^{-3}+9(g_s^V)^2$ of Ref.~\cite{Lee2026} and the experimental input obtained from the PDG~\cite{PDG2026}, $R_\phi^{\mathrm{exp}}=(8.27\pm0.34)\times10^{-3}$. Best-fit points and $1\sigma$, $2\sigma$, and $3\sigma$ regions are shown in Fig.~\ref{fig:fit}. For all three benchmarks, the minimum $\chi^2$ are nearly identical, $\chi^2_{\min}\simeq0.98$, with almost overlapping best-fit points and contours. Taking $|g_u^V|=0.5\times10^{-3}$ as an example, we obtain four degenerate best-fit solutions for $\theta=(g_c^V,g_c^A,g_s^V,g_s^A)$: ($\mathrm{I}$) $(8.32,1.09,-2.62,1.36)$, ($\mathrm{II}$) $(8.32,-1.09,-2.62,-1.36)$, ($\mathrm{III}$) $(-8.32,1.09,2.62,1.36)$, and ($\mathrm{IV}$) $(-8.32,-1.09,2.62,-1.36)$, all in units of $10^{-3}$. Solutions $\mathrm{I}$ and $\mathrm{II}$ correspond to $g_u^V>0$, while solutions $\mathrm{III}$ and $\mathrm{IV}$ correspond to $g_u^V<0$.

Having constrained the couplings through the combined four-channel fit, we further predict the decay ratio for the unmeasured $D^{*+}\to D^+e^+e^-$ channel. For $(g_u^V,g_d^V)=(\pm0.5,\mp2.9)\times10^{-3}$, the four degenerate best-fit solutions given above yield the same result, $R_{D^{*+}}^X\simeq0.5$, compared with the SM prediction $R_{D^{*+}}^{\mathrm{SM}}\simeq6.4\times10^{-3}$. This large enhancement is dominated by the axial-vector contribution $R_A^X$, which far exceeds the corresponding results in the vector coupling scenario listed in Table~\ref{tab:pure_vector}. Scanning the allowed parameter regions, we obtain $R_{D^{*+}}^X\simeq0.35$--$0.65$, $0.26$--$0.75$, and $0.17$--$0.84$ at the $1\sigma$, $2\sigma$, and $3\sigma$ levels, respectively, while even the lowest predicted value is approximately 26 times the SM expectation. The weak dependence on the nuclear benchmarks and the significant enhancement make $D^{*+}\to D^+e^+e^-$ a particularly sensitive channel for testing our theoretical framework.

\section{Conclusion and outlook}

Since the spin and parity of the hypothetical X17 particle remain undetermined, various theoretical interpretations have been proposed. Meson Dalitz decays provide a promising probe of the interaction structure of X17. A previous study has shown that the vector coupling scenario is in tension with the combined constraints from existing Dalitz decay measurements and nuclear transition analyses, thereby motivating us to consider a more general coupling framework. By introducing the axial-vector couplings $g_c^A$ and $g_s^A$, we obtained allowed regions with $|g_{c,s}^A|\sim10^{-3}$, shown in Fig.~\ref{fig:coupling_allowed}, that are compatible with the measured $D^{*0}$, $D_s^{*+}$, $\psi(2S)$, and $\phi$ Dalitz decays. The combined four-channel $\chi^2$ fit, shown in Fig.~\ref{fig:fit}, yields four degenerate best-fit solutions, one of which is $(g_c^V,g_c^A,g_s^V,g_s^A)=(8.32,1.09,-2.62,1.36)\times10^{-3}$, and predicts a significant enhancement of $R^{\rm th}$ in the unmeasured $D^{*+}\to D^+e^+e^-$ decay. The same framework can be extended to the Dalitz decays $B^{*+}\to B^+e^+e^-$, $B^{*0}\to B^0e^+e^-$, $B^*_s\to B_se^+e^-$, and $B^*_c\to B_ce^+e^-$, which could be used to constrain the bottom quark coupling once experimental measurements become available.

A recent study reconsidered the nuclear transition constraints in a framework with both vector and axial-vector couplings~\cite{Fieg2026}. Motivated by this analysis, we take a trial region $(g_u^V,g_d^V,g_u^A,g_d^A)\in([0.17,0.27],[-1.3,-1.1],[-0.058,0.058],[-0.097,0.097])\times10^{-3}$ as input and repeat the combined four-channel fit. Despite allowing nonzero $g_u^A$ and $g_d^A$, the fitted charm and strange quark couplings change by less than about $4\%$, while the minimum value of the combined $\chi^2$ remains around $\chi^2_{\min}\simeq0.98$. This suggests that our results for the charm and strange quark couplings are relatively insensitive to the input $u$ and $d$ quark couplings. However, nonzero $g_d^A$ can notably affect $R^{\mathrm{th}}(D^{*+}\to D^+e^+e^-)$, which is strongly suppressed near $g_d^A\simeq1.4\,g_c^A$. Further experimental measurements of heavy meson Dalitz decays, together with more refined theoretical analyses, will provide essential tests of the coupling structure explored in this work.

\section*{Acknowledgements}
This work was supported by the National Natural Science Foundation of China (NSFC) under Grants Nos. 12375085 and 12575097.


\bibliographystyle{elsarticle-num}
\bibliography{references}






\end{document}